\documentclass[runningheads]{llncs}
\usepackage{graphicx}
\usepackage{amssymb}
\usepackage{amsmath}
\usepackage{comment}
\usepackage{bm}
\usepackage{subfigure}

\begin{document}
	\title{An Accurate Approximation of Resource Request Distributions in Millimeter Wave 3GPP~New Radio Systems}
	\titlerunning{An Approximation of Resource Distributions in mmWave 3GPP NR Systems}
	%
	\author{Roman~Kovalchukov\inst{1}\orcidID{0000-0002-1641-3382} \and
		Dmitri~Moltchanov\inst{1}\orcidID{0000-0003-4007-7187} \and
		Yuliya~Gaidamaka\inst{2,3}\orcidID{0000-0003-2655-4805}\and
		Ekaterina~Bobrikova\inst{2}\orcidID{0000-0002-7704-5827}}
	\authorrunning{R. Kovalchukov et al.}
	%
	\institute{Tampere University, Korkeakoulunkatu 10, Tampere, 33720, Finland \and
		Peoples’ Friendship University of Russia (RUDN University) \\
		6 Miklukho-Maklaya St, Moscow, 117198, Russian Federation \and
		Federal Research Center ``Computer Science and Control''
		of the Russian Academy of Sciences (FRC CSC RAS), 
		44-2 Vavilov St, Moscow, 119333, Russian Federation  \email{rnkovalchukov@sci.pfu.edu.ru}}
	\maketitle              
	\begin{abstract}
		The recently standardized millimeter wave-based 3GPP New Radio technology is expected to become an enabler for both enhanced Mobile Broadband (eMBB) and ultra-reliable low latency communication (URLLC) services specified to future 5G systems. One of the first steps in mathematical modeling of such systems is the characterization of the session resource request probability mass function (pmf) as a function of the channel conditions, cell size, application demands, user location and system parameters including modulation and coding schemes employed at the air interface. Unfortunately, this pmf cannot be expressed via elementary functions. In this paper, we develop an accurate approximation of the sought pmf. First, we show that Normal distribution provides a fairly accurate approximation to the cumulative distribution function (CDF) of the signal-to-noise ratio for communication systems operating in the millimeter frequency band, further allowing evaluating the resource request pmf via error function. We also investigate the impact of shadow fading on the resource request pmf.
	\end{abstract}

	\section{Introduction}
	
	
	
	The future 5G New Radio~(NR) systems are expected to provide three primary services, massive machine-type communications (MTC), enhanced mobile broadband~(eMMB) and ultra-reliable low-latency communications~(URLLC). NR interface operating in the millimeter frequency range is planned to become enabling technology for the latter two services~\cite{ometov2019packet}. The first two phases of NR standardization providing LTE-anchored and standalone NR operations have been completed by 3GPP in December 2017 and August 2018, respectively. The~NR standardization efforts are expected to commence by the end of 2020.
	
	
	
	In addition to inherent advantages of 5G NR related to the use directional antenna radiation and reception patterns and extremely wide bandwidth, the use of millimeter-wave frequency band ($30-100$ GHz) brings unique challenges to system designers, e.g., blockage of propagation between communicating entities path that may lead to abrupt fluctuations of the signal-to-noise ratio~\cite{gapeyenkoICC,moltchanov2019analytical,gapeyenko2017temporal,Samuylov16}. To provide deployment guidelines for 5G NR network operators and evaluate forthcoming technology under a wide variety of prospective scenarios, researchers currently analyze the performance of NR systems in various deployments. The effects of three-dimensional communications scenarios in 5G NR have been addressed in~\cite{kovalchukov2019evaluating}. In~\cite{kovalchukov2018analyzing,gapeyenko2018flexible} the authors have analyzed performance aerial access points operating in the millimeter-wave band. Aiming to improve spectral efficiency and reduce outage probability, the authors~\cite{gapeyenko2018degree} have deeply investigated the effect of multi-connectivity option recently proposed by 3GPP. The upper bound on spectral efficiency in the presence of multi-connectivity has been developed in~\cite{moltchanov2018upper}.

	
	Most of the performance evaluation studies of 5G~NR technology carried out so far concentrated on system aspects characterizing time-averaged user performance using spectral efficiency, achieved rate, and outage probability as the primary metrics of interest. However, the prospective applications of 5G~NR include applications generating bandwidth-greedy non-elastic traffic patterns. Thus, in addition to spatial randomness of users request distributions, performance evaluation models need to capture traffic dynamics as well. Recently, these studies started to appear. In~\cite{petrov2017dynamic}, the authors developed a framework that jointly captures spatial and session-level traffic dynamics in 5G~NR systems in the presence of a 3GPP multi-connectivity option. The authors in~\cite{moltchanov2018improving} proposed a new approach to improve the reliability of the session service process at 5G~NR base stations (BS) using the concept of resource reservation. The effects of both multi-connectivity and bandwidth reservation have been studied in~\cite{kovalchukov2018improved}, where the authors demonstrated that initial selection of NR BS having sufficient amount of resources to handle arriving session provides the positive impact of new and ongoing session drop probabilities. The effect of multi-RAT NR/LTE service process in the street deployment of 5G NR systems has been investigated in~\cite{petrov2018achieving,begishev2018connectivity}. Finally, the joint support of multicast and unicast sessions in 5G~NR systems has been analyzed in~\cite{samuvlov2018performance}.
	
	
	
	Accounting for traffic dynamics at the 5G NR air interfaces requires the joint use of queuing theory and stochastic geometry. Due to the randomness of user locations in the service area as well as propagation and environmental dynamics, the queuing models for this type of analysis need to capture random session resource requirements by \cite{samouylov2016sojourn,naumov2017analysis}. Thus, the critical part in most of the abovementioned studies is a derivation of probability mass function (pmf) of resources required by a session from NR BS. As one may observe, this pmf is the function of multiple system parameters including antenna gains at transmitting and receiving side, emitted power, interference, the randomness of user location within the service area of interest and propagation environments including the density of dynamic blockers. As a result, no closed-form expression is available for the sought pmf. Furthermore, the use of various approximations by the authors to simplify the derivations of the pmf in question restrains potential readers from comparing the reported results across the studies.
	
	
	This study aims to unify the efforts towards an accurate and reliable approximation of pmf of session resource requirements in millimeter-wave 5G NR systems. Using the standardized propagation model, typically assumed coverage of NR BS and random user equipment (UE) distribution we first demonstrate that the signal-to-noise ratio (SNR) perceived at UE can be fairly well approximated by Normal distribution. This fact allows expressing the pmf of session resource requirements in terms of well-known error function drastically reducing the computational efforts. Our numerical results confirm that the proposed model provides an accurate approximation for the session request distribution. Finally, we investigate the effect of shadow fading on the considered pmf.
	
	
	
	The paper is organized as follows. Section~\ref{sect:system} describes the system model. The develop approximation for session resource request pmf in Section \ref{sect:meth}. Numerical illustration of the proposed approximation is provided in Section \ref{sect:numeric}. The last section concludes the paper.

	\section{System Model}\label{sect:system}

	
	\begin{figure}[t]
		\centering
		\includegraphics[width=0.8\textwidth]{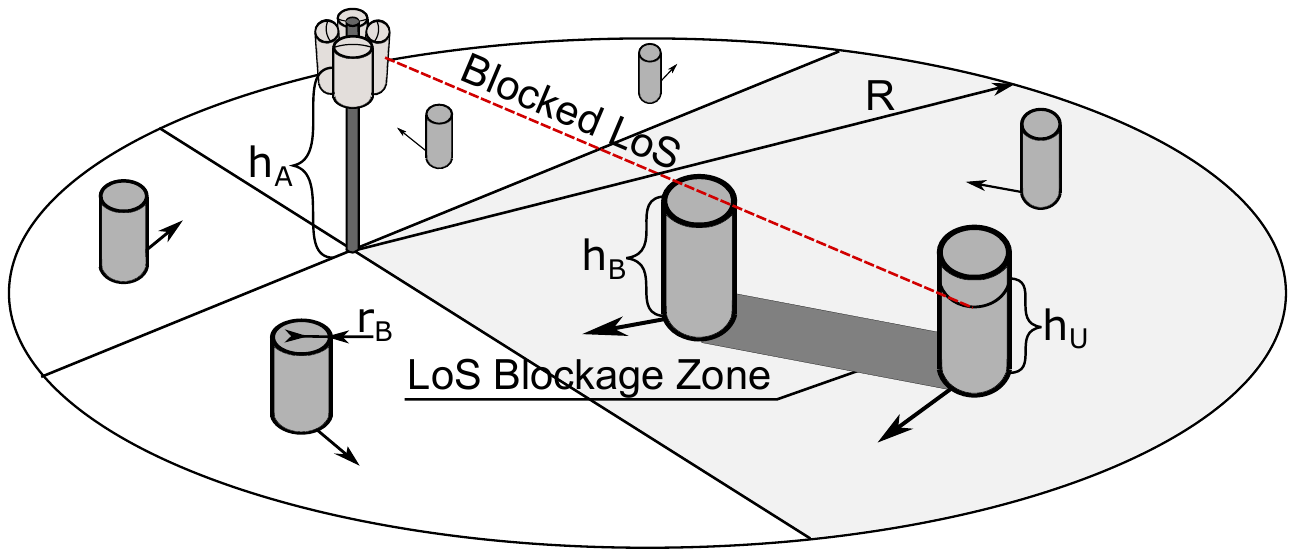}
		\caption{An illustration of the considered 5G NR cellular deployment.}
		\label{fig:deployment}
	\end{figure}  
	
	\subsubsection{Deployment}
	
	
	Fig.~\ref{fig:deployment} illustrates the considered deployment. We assume a single NR BS with a certain coverage area with radius $r_A$ around it. In practice, this is achieved by using several antenna arrays, each service its sector. In what follows, we consider a single sector. The coverage of NR BS is determined by the propagation model specified below, cell-edge outage probability $p_C$, and the set of modulation and coding schemes for 5G NR specified in~\cite{nrmcs}. UE is assumed to be randomly and uniformly distributed in the service area of an NR BS sector. The height of UE and NR BS are assumed to be $h_U$ and $h_A$, respectively.
	
	
	In our scenario, similarly to~\cite{gapeyenko2017temporal}, we also assume dynamic blockage by the mobile crowd around UE. The spatial density of blockers is assumed to be $\lambda_B$. Blockers move around the area according to random direction mobility model (RDM, \cite{nain2005properties}). The flux of blockers across the cell boundary is assumed to be constant; i.e., the density of blockers is homogeneous. Blockers are modeled as cylinders with constant base radius $r_B$ and constant height $h_B$.
	
	\begin{table}[!t]
		\caption{Notation used in the paper.}
		\label{tab:notation}
		\begin{center}
			\vspace{-4mm}
			\begin{tabular}{p{0.2\columnwidth}p{0.71\columnwidth}}
				\hline\hline
				\textbf{Parameter}             & \textbf{Definition}   \\
				\hline\hline
				$f_c$                          & Carrier frequency\\
				\hline
				$\lambda_B$                    & Users density\\
				\hline
				$h_A$                          & NR BS height \\
				\hline
				$r_A$                          &Radius of coverage area\\
				\hline
				$h_U$                          & UE height \\
				\hline
				$h_B$                          & Height of blockers \\
				\hline
				$r_B$                          & Blocker radius \\
				\hline
				$\zeta$                        & Path loss exponent\\
				\hline
				$N_0$                          & Thermal noise\\
				\hline
				$L_B$                          & Loss blockage loss\\
				\hline
				$P_T$                          & NR BS transmit power\\
				\hline
				$C_O$                          & Control channel overhead\\
				\hline
				$C_L$                          & Cable losses\\
				\hline
				$M_I$                          & Interference margin\\
				\hline
				$M_{S,nB},M_{S,B}$             & Shadow fading margins in non-blocked and blocked states\\
				\hline
				$\sigma_{S,B},\sigma_{S,nB}$   & STD of fading in LoS blocked and non-blocked states\\
				\hline
				$N_F$                          & Noise figure\\
				\hline
				$p_C$                          & Cell edge coverage probability\\
				\hline
				$T$                            & Session rate\\
				\hline
				$K_B, K_U$                     & Number of planar antenna elements at NR BS and UE\\
				\hline
				$\omega_B,\omega_U$            & Antenna directivities at NR BS and UE\\
				\hline
				$G_B,G_U$                      & NR BS transmit and UE receive antenna gains\\
				\hline
				$F_X(x),f_{X}(x)$              & CDF and pdf of random variable $X$\\
				\hline
				$S_{\min}$                     & SNR outage threshold \\
				\hline
				$S_i,S$                        & SNR with NR BS $i$ and overall SNR\\
				\hline
				$p_{B,i}(x),p_{B,i}$           & Distance-dependent/ independent blockage probabilities \\
				\hline
				$s_j$                          & SNR margins\\
				\hline
				$m_j$                          & Probability of choosing MCS $j$\\
				\hline
				$e_j$                          & Spectral efficiency of MCS $j$\\
				\hline
				$\text{erfc}(\cdot)$           & Complementary error function\\
				\hline\hline
			\end{tabular}
			\vspace{-4mm}
		\end{center}
		\vspace{-4mm}
	\end{table}
	
	\subsubsection{Propagation Model and SNR}
	
	
	
	The SNR at UE can be written as
	\begin{align}\label{eqn:genericProp}
	P_{R}(y)=\frac{P_{T}G_{B}G_{U}}{L_{dB}(y)N_0C_OC_LM_IN_FM_S},
	\end{align}
	where $P_{T}$ is the NR BS transmit power, $G_B$ and $G_{U}$ are the antenna gains at the NR BS and UE sides, respectively, $y$ is the three-dimensional (3D) distance between the UE and the NR BS, $L_{dB}(y)$ is the propagation loss in decibels, $C_O$~is the control channel overhead , $C_L$ is the cable losses, $M_I$ is the interference margin, $N_F$ is the noise figure, $M_S$ is the shadow fading margin.
	
	
	We capture interference from adjacent NR BSs via interference margin $M_I$. For a given NR BS deployment density, one may estimate the interference margin using stochastic geometry-based models ~\cite{kovalchukov2019evaluating,kovalchukov2018analyzing,petrov2017interference}.  The effect of shadow fading is accounted for using shadow fading margins, $M_{S,B}$, $M_{S,nB}$ for LoS blocked, and non-blocked states provided in~\cite{standard38901}.
	
	
	The LoS path between the UE and the NR BS might be temporarily occluded by moving users. Depending on the current link state (LoS blocked or non-blocked) as well as the distance between the NR BS and the UE, the running session employs an appropriate MCS specified in TR 38.211 to maintain reliable data transmission~\cite{nrmcs}. We also utilize the 3GPP urban micro (UMi) street canyon model specified in TR 38.901 with blockage enhancements that provide path loss for a certain separation distance with and without blockage~\cite{standard38901}. Particularly, the path loss is
	\begin{align}
	\hspace{-2mm}L_{dB}(y)=
	\begin{cases}
	32.4 + 21\log(y) + 20\log{f_c},\,\:\text{non-blocked},\\
	52.4 + 21\log(y) + 20\log{f_c},\,\:\text{blocked},
	\end{cases}
	\end{align}
	where $y$ is the 3D distance, $f_c$ is the carrier frequency in GHz. 

	\subsubsection{Session Resource Requirements}\label{sect:pmf}
	
	
	We assume that the session requires constant bitrate $R$. Technically, to determine pmf of resources required from NR BS to serve a session with bitrate $R$, we have to know the CQI and MCS values as well as SNR to CQI mapping. As these parameters are usually vendor-specific, in our study, we use MCS mappings from~\cite{fan2011mcs} provided in Table~\ref{tab:mapping}.
	
	\begin{table}[!t]
		\centering
		\caption{CQI, MCS and SNR mapping for 3GPP NR.}
		\label{tab:mapping}
		\begin{center}
			\vspace{-4mm}
			\begin{tabular}{p{0.1\columnwidth}p{0.28\columnwidth}p{0.32\columnwidth}p{0.2\columnwidth}}
				\hline\hline
				\textbf{CQI}            & \textbf{MCS}  & \textbf{Spectral efficiency} & \textbf{SNR in dB} \\
				\hline\hline
				0                         &  out of range\\
				\hline
				1                              & QPSK, 78/1024 &0.15237 &-9.478 \\
				\hline
				2                            & QPSK, 120/1024 &0.2344 &-6.658\\
				\hline
				3                            & QPSK, 193/1024 &0.377 &-4.098 \\
				\hline
				4                            & QPSK, 308/1024 &0.6016 &-1.798\\
				\hline
				5                            & QPSK, 449/1024 &0.877 &0.399 \\
				\hline
				6                            & QPSK, 602/1024 &1.1758 &2.424\\
				\hline
				7                            & 16QAM, 378/1024 &1.4766 &4.489\\
				\hline
				8                            & 16QAM, 490/1024 &1.9141 &6.367 \\
				\hline
				9                            & 16QAM, 616/1024 &2.4063 &8.456 \\
				\hline
				10                            & 16QAM, 466/1024 &2.7305 &10.266\\
				\hline
				11                            & 16QAM, 567/1024 &3.3223 &12.218\\
				\hline
				12                            & 16QAM, 666/1024 &3.9023 &14.122\\
				\hline
				13                            & 16QAM, 772/1024 &4.5234 &15.849\\
				\hline
				14                            & 16QAM, 873/1024 &5.1152 & 17.786\\
				\hline
				15                            & 16QAM, 948/1024 &5.5547 & 19.809\\
				\hline\hline
			\end{tabular}
			\vspace{-4mm}
		\end{center}
		\vspace{-4mm}
	\end{table}
	
	
	Denote by $s_j$, $j= 1,2,..,K$, the SNR margins of the NR MCS schemes, where $K$ is the MCS number, and by $m_j$ the probability that the UE session is assigned to MCS $j$. We have
	\begin{align}\label{eqn:mi}
	m_j=\Pr\{s_j<s<s_{j+1}\}=W_S(s_{j+1})-W_S(s_j),
	\end{align}
	where $F_S(x)$ is the CDF of SNR $S$.
	
	Once $m_j$, $j= 1,2,\dots,K$, are available, the probabilities that user will request $i$ resources for a session with rate $R$ is provided as
	\begin{align}\label{eqn:pi}
	p_i=\sum_{\forall{j}:e_j\in \left[\frac{R}{i W_{\text{PRB}}},\frac{R}{(i-1) W_{\text{PRB}}}\right)}m_j,
	\end{align}
	where $e_j$ is a spectral efficiency of $j$-th CQI and $W_{\text{PRB}}$ is a bandwidth of the primary resource block (PRB). 

	\section{The Proposed Methodology}\label{sect:meth}
	
	In this section, we develop the approximation for pmf of session resource requirements in the considered scenario. First, we determine the maximum coverage area of NR BS such that cell edge UE experiences no more than a fraction of time $p_C$ in an outage. Next, we develop approximation for SNR that simultaneously accounts for random UE location in the service area and shadow fading. Finally, we derive the pmf of the session resource requirements.
	
	\subsection{NR BS Coverage}
	
	
	
	We first determine maximum coverage of the deployment area, $r_A$, such that no UEs experience outage with any of NR BS located on the circumference. Let $S_{\min}$ be the SNR outage threshold, i.e., $S_{\min}$ is the lower bound of the SNR range corresponding to the lowest MCS~\cite{nrmcs}. Using the propagation model for LoS blockage state, we have the following relation
	\begin{align}\label{eqn:smin}
	S_{\min}=\frac{P_T G_B G_U}{N_0C_OC_LM_IN_FM_{S,B}}(r_A+[h_A-h_U]^2)^{-\zeta/2},
	\end{align}
	where $\zeta$ is the path loss exponent, $h_A$ and $h_U$ are the heights of NR BS and UE, $P_T$ is the NR BS transmit power, $G_B$ and $G_U$ are the NR BS transmit and the UE receive antenna gains, $N_0$ is the thermal noise, $C_O$ is the control channel overhead, $C_L$ is the cable losses, $M_I$ is the interference margin, $N_F$ is the noise figure, $M_{S,B}$ is the fading margin in LoS blocked state.
	
	Solving \eqref{eqn:smin} with respect to $r_A$, we obtain
	\begin{align}
	r_A=\sqrt{\left(\frac{P_TG_BG_U}{N_0C_OC_LM_IN_FM_{S,B}S_{\min}}\right)^{\zeta/2}+(h_A - h_U)^2},
	\end{align}
	where $M_{S,B}$ is computed as follows
	\begin{align}
	M_{S,B}=\sqrt{2}\sigma_{S,B}\text{erfc}^{-1}(2p_C),
	\end{align}
	where $\text{erfc}^{-1}(\cdot)$ is the inverse complementary error function, $p_C$ is the cell edge coverage probability, and $\sigma_{S,B}$ is standard deviation (STD) of shadow fading distribution for LoS blocked state, which is provided in~\cite{standard38901}.
	
	\subsection{SNR CDF Approximation}
	
	
	We now proceed deriving SNR CDF. Observe that in the considered model, the randomness of SNR is due to two factors, UE location, and shadow fading. For accounting for shadow fading, that follows Lognormal distribution (i.e., Normal distribution in the decibel scale), it is easier to operate in decibel~scale. 
	
	To derive SNR CDF approximation we first obtain CDF of the 3D distance between UE and NR BS assuming that the position of UE is uniformly distributed within the coverage zone. Recall, that the 2D distance is distributed according to probability density function (pdf) $w_{R}(x)=2x/2r_A$~\cite{distanceMoltchanov}. Now, 3D distance can be expressed as a function of 2D distance using $\phi_D(r)=\sqrt{(h_A - h_U)^2 + r^2}$. The 3D distance can be found using the random variable (RV) transformation technique~\cite{ross}. Particularly, recall that pdf of a RV $Y$, $w(y)$, expressed as function $y=\phi(x)$ of another RV $X$ with pdf $f(x)$ is
	\begin{align}\label{eqn:rvTrans}
	w(y)=\sum_{\forall{i}}f(\psi_i(y))\left|\frac{d\psi_i{'}(y)}{dy}\right|,
	\end{align}
	where $x=\psi_i(y)=\phi^{-1}(x)$ is the inverse functions.
	
	Substituting $w_r(x)$ and $\phi_r(x)$ into~\eqref{eqn:rvTrans} we arrive at
	\begin{align}
	W_{d}(x)=
	\begin{cases}
	1&\sqrt {d_E^2 + h_A^2 - 2{h_A}{h_U} + {h_U}^2} \le x,\\
	\frac{{2{h_A}{h_U} -h_A^2+ h_U^2 + {x^2}}}{{d_E^2}},& {h_A} - {h_U}< x < \sqrt {d_E^2 + h_A^2 - 2{h_A}{h_U} + h_U^2},\\
	0&\text{elsewhere.}\\
	\end{cases}
	\end{align}

	
	The inverse of the SNR in decibel without shadow fading can be found using the same technique, where SNR in decibels is a function of 3D distance $\phi_{SNR,dB}(d)=10 \log _{10}\left(Ad^{-\zeta}\right)$, where $d$ is a 3D distance between UE and NR BS, $\zeta$ is a pathloss exponent and $A$ is a term representing all gains and losses except propagation losses and fluctuations due to shadow fading. Substituting $\phi_{SNR,dB}(x)$ and $W_d(x)$ into \eqref{eqn:rvTrans} we get SNR CDF:
	\begin{align}\label{eqn:deriv}
	{W_{S^{dB}}}(x) =
	\begin{cases}
	1 - \frac{{{{10}^{ - \frac{x}{{5\zeta }}}}{A^{2/\Gamma }} - {{\left( {{h_A} - {h_U}} \right)}^2}}}{{d_E^2}} &x>10\log 10\left[ {A{{B}^{ - \zeta /2}}} \right],\\
	0&\text{elsewhere,}\\
	\end{cases}
	\end{align}
	where $B={d_E^2 + {{\left( {{h_A} - {h_U}} \right)}^2}}$.
	
	Now, recalling that shadow fading is characterized by Lognormal distribution in linear scale leading to Normal distribution in the decibel scale, the random variable specifying the SNR distribution can be written as
	\begin{align}
	{S_{SF}} = {S^{dB}} + Norm(0,{\sigma _{SF}}).
	\end{align}
	
	Finally, we determine the SNR CDF as a convolution of $W_{S}(y)$ and the probability density function of a normal distribution with zero mean and standard deviation $\sigma$, i.e.,
	\begin{align}
	{W_{{S_{SF}}}}(y) = \int_{ - \infty }^\infty  {{W_S}\left( {y + u} \right)} \frac{{{e^{ - \frac{{{u^2}}}{{2{\sigma ^2}}}}}}}{{\sqrt {2\pi } \sigma }}du.
	\end{align}
	
	Unfortunately, the latter cannot be evaluated in closed-form by using random variables transformation technique but can be represented in terms of error functions as follows:
	\begin{align}\label{eqn:sfsnr}
	&{W_{{S_{SF}}}}(x) = \frac{1}{{2d_E^2}} \left[ {{A^{2/\gamma }}{{10}^{ - \frac{x}{{5\zeta }}}}{e^{\frac{{{\sigma ^2}{{\log }^2}(10)}}{{50{\gamma ^2}}}}}} \right.\nonumber\\
	&\left[ {{\rm{erf}}\left( {\frac{{50\zeta \log A - 25{\gamma ^2}\log B + {\sigma ^2}{{\log }^2}10 - 5\zeta x\log 10}}{{5\sqrt 2 \gamma \sigma \log (10)}}} \right) - } \right.\nonumber\\
	&\left. { - {\rm{erf}}\left( {\frac{{50\zeta (\log A - \gamma \log ({h_A} - {h_U})) + \sigma _S^2{{\log }^2}10 - 5\zeta x\log (10)}}{{5\sqrt 2 \gamma \sigma \log (10)}}} \right)} +\right]\nonumber\\
	&+ \left( {d_E^2 + {{({h_A} - {h_U})}^2}} \right) {\rm{erf}}\left( {\frac{{ - 10\log A + 5\zeta \log B + x\log 10}}{{\sqrt 2 \sigma \log (10)}}} \right)- {{({h_A} - {h_U})}^2}\times{}\nonumber\\
	&\left. { \times{}{\rm{erf}}\left( {\frac{{\sqrt 2 ( - 10\log A + 10\zeta \log ({h_A} - {h_U}) + x\log 10)}}{{\sigma \log 100}}} \right) + d_E^2} \right],
	\end{align}
	where $B={d_E^2 + {{\left( {{h_A} - {h_U}} \right)}^2}}$, $\text{erf}(\cdot)$ is the error function.
	
	Including the blockage induced losses $L_B$ into $A$ and using $\sigma_{S,B}$ and $\sigma_{S,nB}$ into (\ref{eqn:sfsnr}), we can obtain two SNR CDFs $W_{S_{nB}}$ and $W_{S_{B}}$ for non-blocked LoS and blocked LoS conditions.
	
	
To determine SNR $S_i$, we also need the blockage probability.
Observe that with the specified RDM mobility model the fraction of time UE located at the 2D distance $x$ from NR BS is in blocked conditions coincides with the blockage probability provided in~\cite{gerasimenko2019capacity},
	\begin{align}
	p_{B}(x)=1-e^{-2\lambda_Br_B\left[x\frac{h_{B}-h_{U}}{h_{A}-h_{U}}+r_B\right]}.
	\end{align}
	leading to the following weighted blockage probability with the NR BS
	\begin{align}
	p_{B}=\int_{0}^{r_A}p_{B}(x)w_{D}(x)dx.
	\end{align}
	
	The final result for SNR CDF accounting for shadow fading and blockage is
	\begin{align}\label{eqn:fullsnr}
	W_S(x)=P_BW_{S_{B}(x)}+\left(1-P_B\right)W_{S_{nB}(x)}.
	\end{align}
	
	Once SNR CDF is obtained, session resource requirements pmf can be obtained using (\ref{eqn:mi}) and (\ref{eqn:pi}). Observe that it is expressed in terms of error functions. 

	\section{Numerical Results}\label{sect:numeric}

	In this section, we report our numerical results. We first demonstrate that the resulting SNR CDF closely follow Normal distribution. Then, we proceed illustrating the effect of shadow fading on session resource requirements pmf. System parameters used in this section is shown in Fig.~\ref{tab:parameters}.
	
	\begin{table}[t!]
		\centering
		\caption{System parameters.}
		\label{tab:parameters}
		\vspace{-4mm}
		\begin{center}
			\bgroup
			\def\arraystretch{1.1}%
			\footnotesize 
			\begin{tabular}{p{0.6\columnwidth}p{0.3\columnwidth}}
				\hline\hline
				\textbf{Parameter} & \textbf{Value} \\
				\hline\hline
				Carrier frequency, $f_c$                & 28 GHz\\
				\hline
				Transmit power, $P_T$                & 23 dBm\\
				\hline
				UE receive antenna gain $G_U$        & 5.57 dBi\\
				\hline
				NR BS transmit  gain $G_B$        & 20.58 dBi\\
				\hline
				LoS blockage loss, $L_B$            & 20 dB\\
				\hline
				NR BS height, $h_A$                & 4 m\\
				\hline
				UE height, $h_U$                    & 1.5 m\\
				\hline
				Blocker height, $h_B$                & 1.7 m\\
				\hline
				Blocker radius, $r_B$                & 0.3 m\\
				\hline
				User density, $\lambda_B$          & 0.2 users/m$^2$\\
				\hline
				Session rate, $R$                    & 2 Mbps\\
				\hline
				Control channel overhead, $C_O$        & 1 dB\\
				\hline
				Cable losses, $C_L$        & 2 dB\\
				\hline
				Interference margin, $M_I$        & 3 dB\\
				\hline
				Thermal noise, $N_0$        & -174 dBm/Hz \\
				\hline
				Noise figure, $N_F$        & 7 dB\\
				\hline
				Noise figure, $W_{PRB}$        & 1.44 Mhz\\
				\hline
				Min SNR, $S_{min}$        & -9.478 dB\\
				\hline
				Cell edge coverage probability, $p_C$        & 0.01, 0.05, 0.1\\
				\hline
				Radius of coverage area, $r_A$        & 65, 119, 165 m\\
				\hline
				Standard deviation of shadow fading, $\sigma_{S,B},\sigma_{S,nB}$        & 4, 8.2 dB\\
				\hline\hline
			\end{tabular}
			\egroup    \vspace{-4mm}
		\end{center}
		\vspace{-4mm}
	\end{table}

	\subsection{SNR Approximation}\label{sect:snr_cdf}
	
	
	
\begin{figure}[t]
		\centering
		\includegraphics[width=1.0\textwidth]{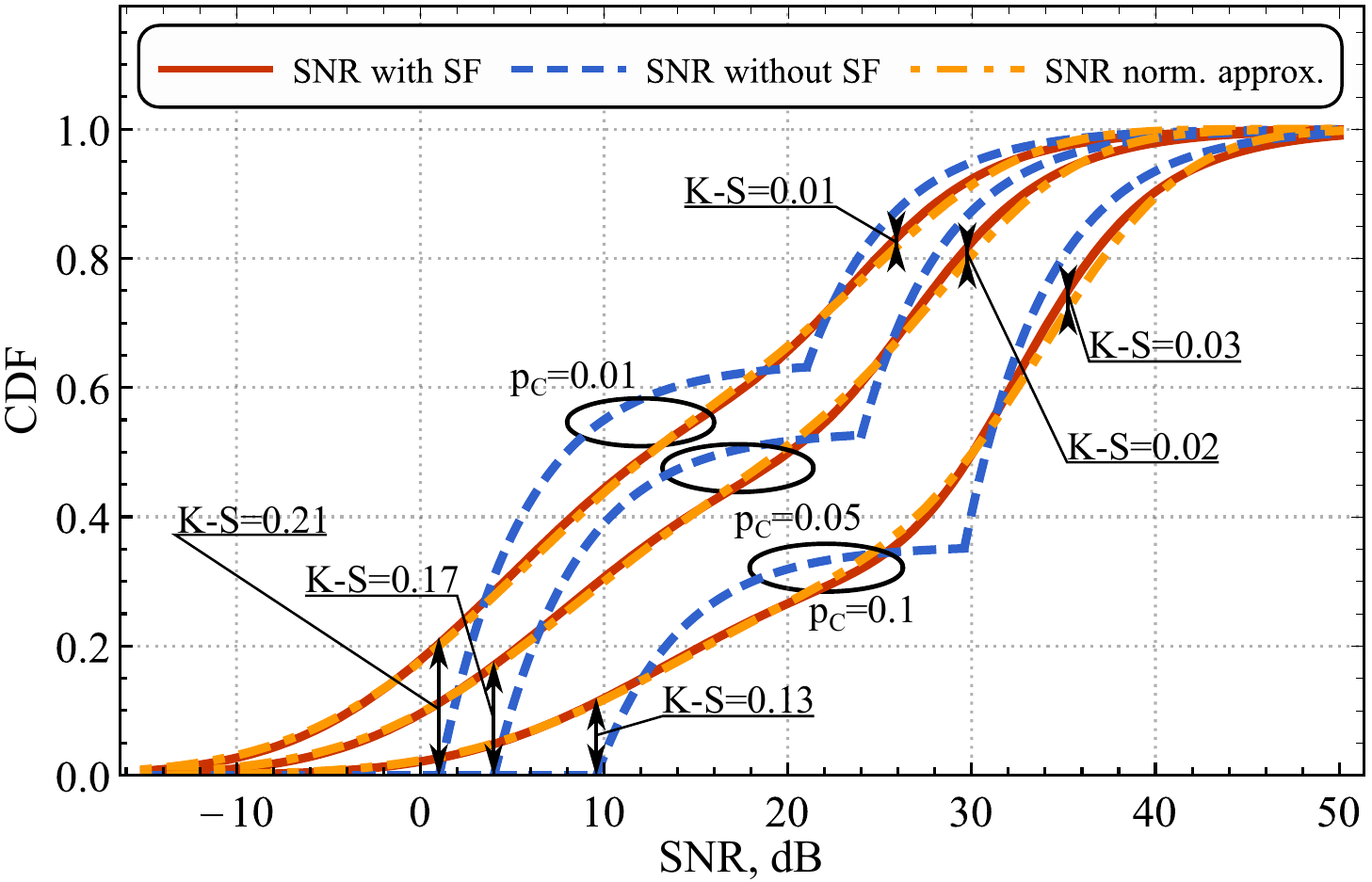}
		\caption{SNR CDF with/without shadow fading and their approximations.}
		\label{fig:snr_cdf}
		\vspace{-4mm}
	\end{figure}

	
	We first start assessing the proposed Normal approximation to SNR CDF. Fig.~\ref{fig:snr_cdf} illustrates SNR CDFs with and without shadow fading as well as their approximations by a weighted sum of two Normal distributions, similar to \eqref{eqn:fullsnr}, with parameters $\mu=E[S^{dB}]$ and $\sigma=\sqrt{\sigma_{SF,\cdot}^{2}+\sigma_{S^{dB}}^{2}}$ for LoS and nLoS cases. To assess the closeness of original and approximating distributions, we use the notion of statistical distance. Particularly, we apply the Kolmogorov Statistic (K-S) also shown in Fig.~\ref{fig:snr_cdf}. Analyzing the resulting values of K-S statistic, the proposed approximation is extremely close to the original SNR CDF with shadow fading. Furthermore, as one may observe, the approximation becomes better standard deviation of shadow fading increases, e.g., in nLoS case.
	
	Note that the Gaussian distribution of shadow fading partially explains the suitability of Normal approximation and partially due to other random effects involved in SNR CDF, e.g., blockage, random UE location.
	
	The second critical observation is that SNR CDF without shadow fading, also shown in Fig.~\ref{fig:snr_cdf} drastically deviates from the one accounting for this effect. Thus, excluding the effects of shadow fading from the model characterizing resource requirements pmf may lead to drastic errors in the predicted system and user performance metrics. Recalling the non-linear mapping of SNR into MCS schemes dropping the effects of shadow fading leads to overly optimistic results.
	
	\subsection{Resource Request Approximation}\label{sect:resource_pmf}
	
	
	\begin{figure}[b!]
		\vspace{-4mm}
		\centering
		\subfigure[{SF vs. no SF}]
		{
			\includegraphics[width=0.47\textwidth]{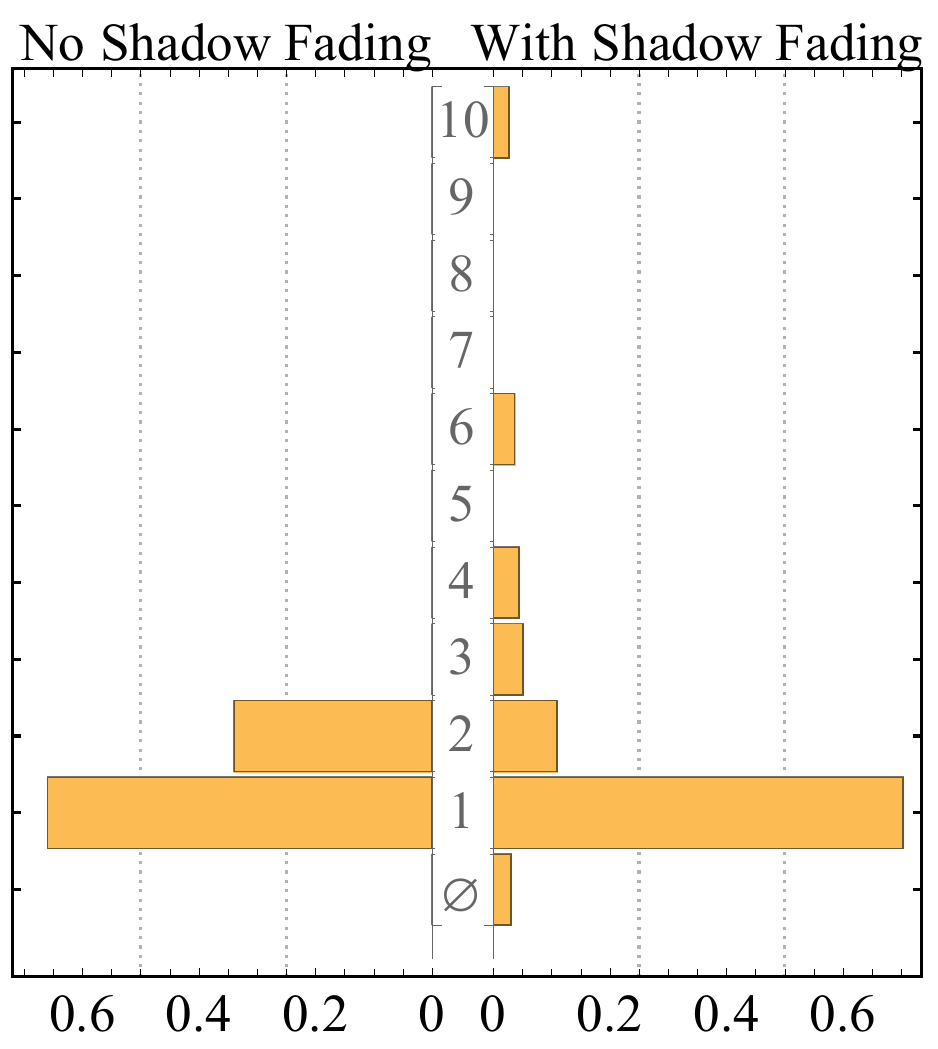}
			\label{fig:pmf1}
		}~
		\subfigure[{SF vs. approximation}]
		{
			\includegraphics[width=0.47\textwidth]{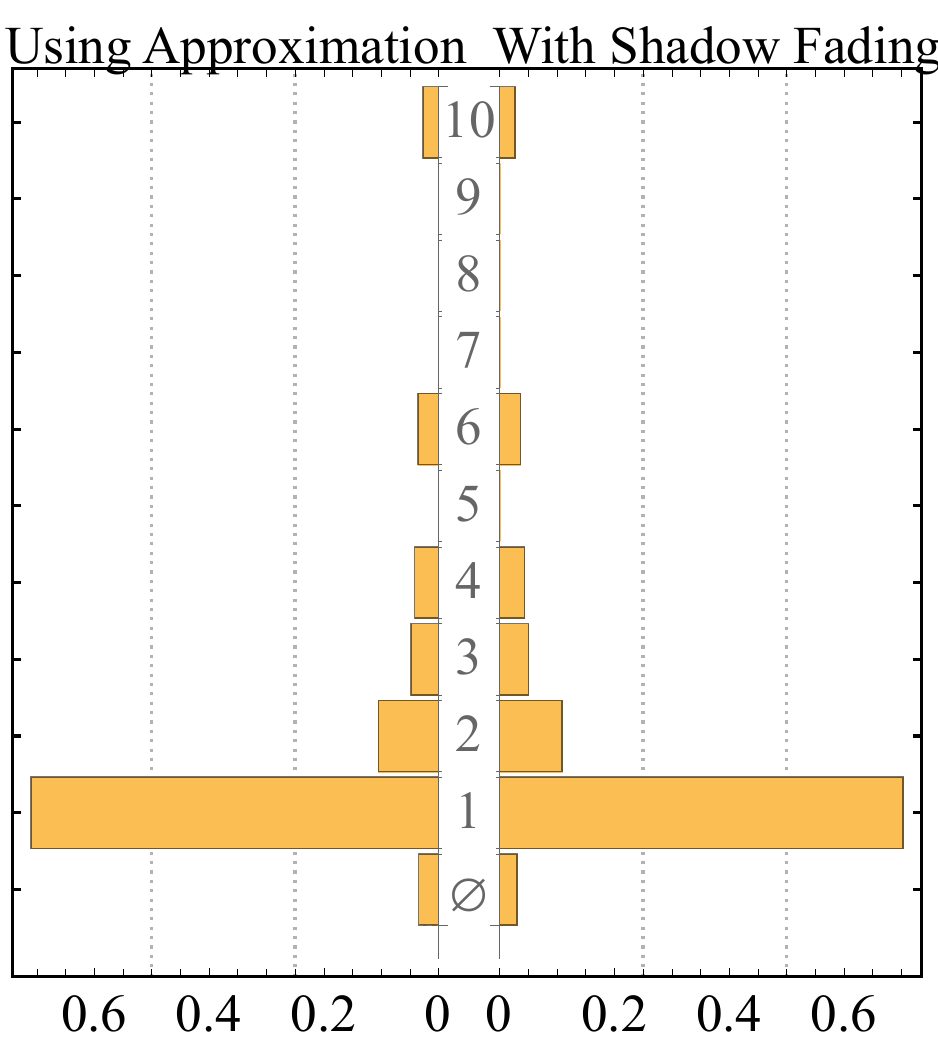}
			\label{fig:pmf2}
		}
		\caption{Comparison of pmfs of resource requirements for session rate 2 Mbps.}
		\label{fig:pmf2Mbps}
	\end{figure}

	\begin{figure}[t!]
		\centering
		\subfigure[{SF vs. no SF}]
		{
			\includegraphics[width=0.47\textwidth]{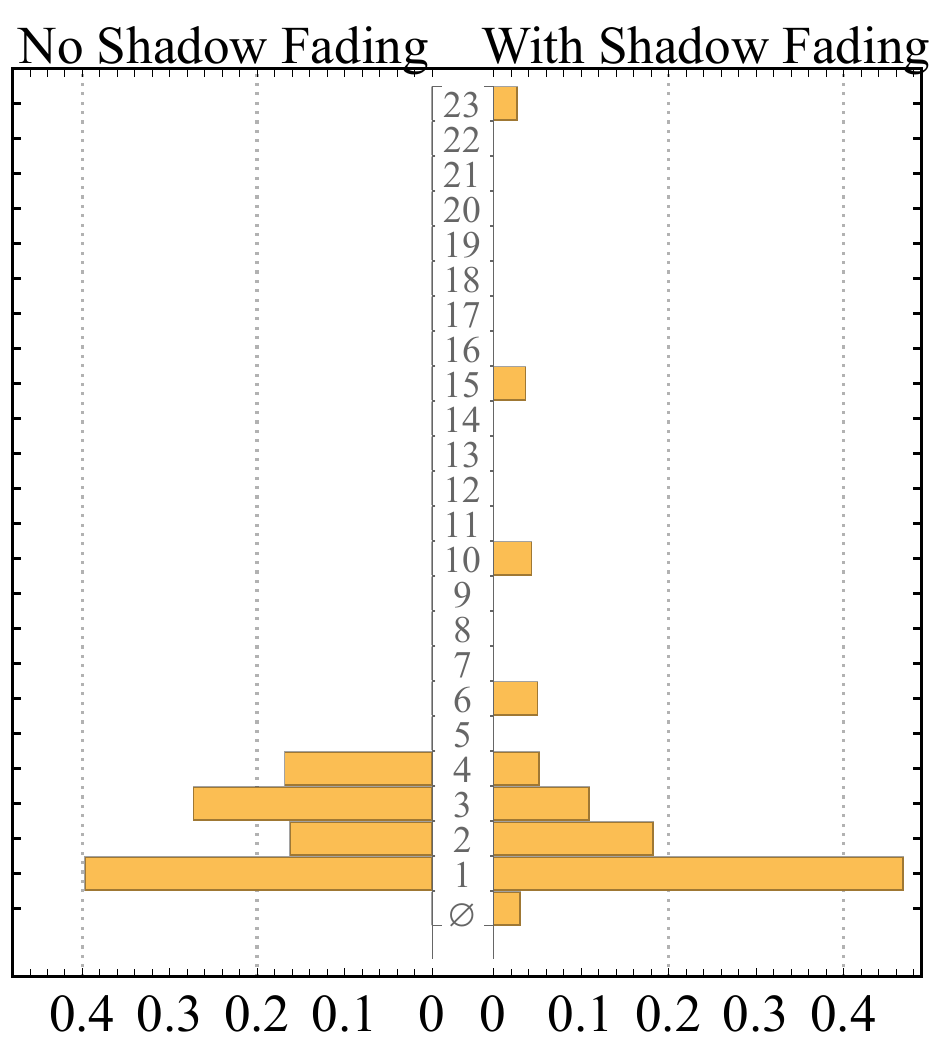}
			\label{fig:pmf3}
		}~
		\subfigure[{SF vs. approximation}]
		{
			\includegraphics[width=0.47\textwidth]{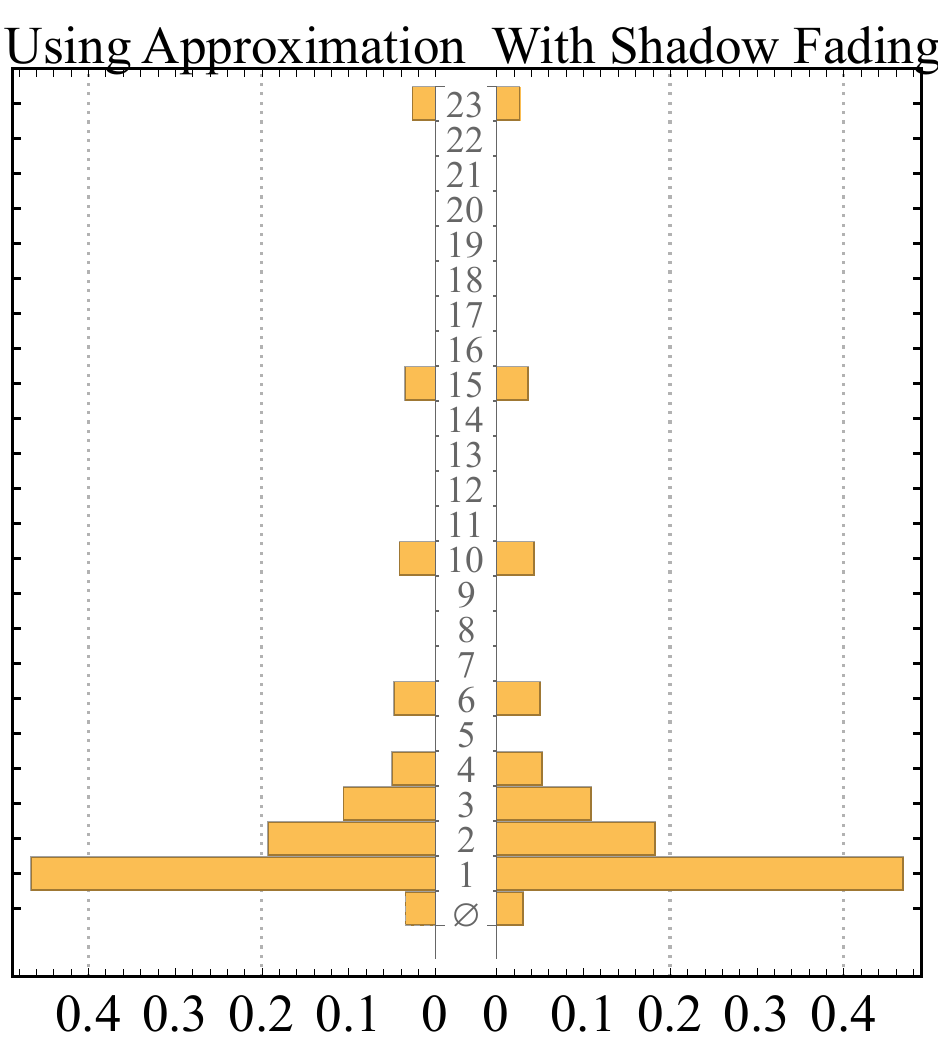}
			\label{fig:pmf4}
		}
		\caption{Comparison of pmfs of resource requirements for session rate 5 Mbps.}
		\label{fig:pmf5Mbps}
		\vspace{-2mm}
	\end{figure}
	
	
	
	We now proceed highlighting the effect of shadow fading on resource request pmf. Observe that its effect manifests itself in two ways: (i) it affects the coverage area of NR BS, $r_A$, and (ii) once $r_A$ is determined the shadow fading affects the number of requested resources directly by introducing another source of uncertainty in addition to distance-induced path losses.
	
	The comparison of resource requirements pmfs with and without shadow fading is provided in Fig.~\ref{fig:pmf2Mbps} for $R=2$ Mbps and in Fig.~\ref{fig:pmf5Mbps} for $R=5$ Mbps. In both cases, the fraction of outage time for cell-edge UE in nLoS state is set to $p_C=0.1$. As one may observe, the difference is rather significant and may drastically affect the absolute values of the performance metrics of interest in 5G NR system analysis. Note that not only the form of the distribution changes but its moments as well. The mean resource requirements with shadow fading taken into account are now higher compared to the model without this effect. 
	
	Table~\ref{tab:res} provides the mean and standard deviation of SNR with and without shadow fading and the proposed approximation for SNR using Normal distribution. It also illustrates the mean and standard deviation of resulting session resource requirements, including original pmf and its approximation. 
	
	
	\begin{table}[t!]
		\centering
		\caption{Summary of approximations for $R=5$ Mbps.}
		\label{tab:res}
		\begin{center}
			\bgroup
			\vspace{-4mm}
			\def\arraystretch{1.1}%
			\footnotesize 
			\begin{tabular}{p{0.55\columnwidth}p{0.13\columnwidth}p{0.13\columnwidth}p{0.1\columnwidth}}
				\hline\hline
				$\bm{p_C}$ & \textbf{0.01} & \textbf{0.05} & \textbf{0.1} \\ 
				\hline\hline
				Mean SNR & 27.016 & 17.8982 & 12.7958 \\ 
				\hline 
				Mean SNR no SF & 27.016 & 17.8982 & 12.7958 \\ 
				\hline 
				Mean SNR Approximation & 27.016 & 17.8982 & 12.7958 \\ 
				\hline\hline
				STD SNR & 12.0514 & 12.7718 & 12.7065 \\ 
				\hline 
				STD SNR no SF & 10.538 & 10.9461 & 10.6159 \\ 
				\hline 
				STD SNR Approximation & 12.0513 & 12.7718 & 12.7065 \\ 
				\hline\hline
				Mean Resource Requirement & 1.42256 & 2.37408 & 3.27115 \\ 
				\hline 
				Mean Resource Requirement no SF & 1.22362 & 1.69017 & 2.21262 \\ 
				\hline 
				Mean Resource Requirement Approximation & 1.43419 & 2.35948 & 3.20999 \\ 
				\hline\hline
				SDT Resource Requirement & 2.71887 & 11.7379 & 20.8665 \\ 
				\hline 
				STD Resource Requirement no SF & 0.17362 & 0.64415 & 1.29765 \\ 
				\hline 
				STD Resource Requirement Approximation & 3.03627 & 11.9472 & 20.5599 \\ 
				\hline \hline
			\end{tabular}
			\egroup
		\end{center}
		\vspace{-5mm}
	\end{table}

	\section{Conclusion}

	Characterizing session request requirements is an essential step in assessing user-level performance provided by forthcoming 5G NR systems. In this study, to derive pmf of resources requested by a session from NR BS we have proposed a unified methodology that accounts for 3GPP propagation model, random location of UE within the coverage area of NR BS and other environmental impairments including shadow fading and blockage of LoS path between UE and NR BS. We further demonstrated that in the presence of shadow fading SNR CDF could be well approximated by Normal distribution that allows expressing session resource requirements pdf in terms of error functions. The proposed methodology can be extended to the case of random requested rate $R$.
	
	
	


	\bibliographystyle{ieeetr}
	\bibliography{refs}
	
\end{document}